\documentclass[12pt]{article}
\pdfoutput=1
\usepackage{graphicx}
\usepackage{hyperref}
\usepackage{amsmath,amssymb}
\DeclareGraphicsExtensions{.jpg}

\def\be{\begin{equation}}
\def\ee{\end{equation}}
\def\bear{\begin{eqnarray}}
\def\eear{\end{eqnarray}}
\def\rc{\nonumber\\ }
\def\RR{{\rm I\kern-1.6pt {\rm R}}}

\newcommand{\tret}{{t_{\mbox{\scriptsize ret}}}}

\begin{document}

\pagestyle{empty}

\begin{center}
{\LARGE{\bf Accelerated detectors 
and worldsheet horizons in AdS/CFT}}
\end{center}
\vskip 10pt
\begin{center}
{\large 
Mariano Chernicoff and 
Angel Paredes}
\end{center}
\vskip 10pt
\begin{center}
\textit{Dept. de F\'\i sica Fonamental and
Institut de Ci\`encies del Cosmos (ICC), Universitat de Barcelona, Marti i Franqu\`es, 1,
E-08028, Barcelona, Spain.
}
{\small  mchernicoff@ub.edu, aparedes@ffn.ub.es}
\end{center}

\vspace{15pt}

\begin{center}
\textbf{Abstract}
\end{center}

\vspace{4pt}{\small \noindent 
We use the $AdS$/CFT correspondence to discuss the response of an accelerated observer to the
quantum vacuum fluctuations.
In particular, we study heavy quarks probing a strongly coupled CFT
by analysing strings moving in $AdS$.
We propose that, in this context, a non-trivial detection rate is associated to
the existence of a worldsheet horizon and we find an Unruh-like expression for the
worldsheet temperature.
Finally, by examining a rotating string in global $AdS$ we find that there is a transition between string embeddings with and without worldsheet horizon. The dual picture
corresponds to having non-trivial or trivial interaction with the quantum vacuum respectively.  This is in qualitative agreement with results of Davies et al.}
\vfill

\newpage

\setcounter{page}{1}
\pagestyle{plain}

\tableofcontents

\section{Introduction and summary of results}

A remarkable feature of quantum field theory is that a detector undergoing non-inertial
motion might measure excitations related to fluctuations of the quantum vacuum. Roughly 
speaking, it detects ``particles" --- or as sometimes said, it {\it clicks} --- even if it is moving in vacuum. The first and
 most studied
example is the Unruh effect \cite{Unruh:1976db} (for a recent review, see
\cite{Crispino:2007eb}): an object in uniform proper acceleration 
--- constant in modulus and direction --- within Minkowski space sees vacuum fluctuations as a
thermal bath at temperature
\be
T_U = \frac{A}{2\pi}\,\frac{\hbar}{k_B c}.
\label{unruh}
\ee
In the following, we will set $\hbar=c=k_B=1$. 
There are different ways of understanding this phenomenon. First, one can realize
that the definition of a particle in quantum field theory is not unique \cite{Fulling:1972md}.
The natural quantization with respect to different time-like Killing vectors is different and
Bogolubov transformations relate annihilation and creation operators in the two frames. 
If positive and negative norm states mix non-trivially, the vacuum defined in
the different frames (state annihilated by all annihilation operators) can be distinct.
In such case, the vacuum associated to a uniformly accelerated trajectory is the Fulling vacuum whereas
the accelerated particle sees the Minkowski vacuum as a thermally populated state.
Second, one can define an Unruh-DeWitt detector
\cite{Unruh:1976db,dewitt,birrelldavies}, in which a scalar field (in its vacuum state) living in space-time
is weakly coupled to a detector in motion.
Self-correlations 
of the scalar field on the trajectory cause the detector to click. Third, 
changing to space-time coordinates in which the accelerated observer is static, the
Minkowski metric becomes the Rindler metric and presents an acceleration horizon. In this case the thermal spectrum is associated to 
Hawking emission coming from the horizon.
These three arguments --- non-trivial Bogolubov transformation, Unruh-DeWitt detector, existence
of Rindler horizon --- lead to equivalent conclusions for the uniformly accelerated particle.
However, results differ for more general motions, as has been remarked and explained in the
literature, see for instance \cite{Padmanabhan:1983ub,Sriramkumar:1999nw}.
This emphasizes that the arguments above refer to different aspects of the physics that cannot be
identified in general.

The simplest example of this seeming disagreement is that of 
circular motion, as first 
discussed by Letaw and Pfautsch \cite{Letaw:1979wy}. An Unruh-DeWitt detector was shown to measure
a (non-thermal) spectrum of excitations\footnote{Circular motion is a rather interesting case since electrons in storage
rings have been proposed as possible experimental ``thermometers" for an acceleration temperature
\cite{Bell:1982qr}. 
Even if for circular motion
the spectrum is not thermal, it is possible
to define an energy dependent temperature, namely, relative populations of states of different energy
will not depend only on the modulus of acceleration but also on the energy difference in a non-trivial way.
See also \cite{Akhmedov} for related work in which the relation to the experimentally measured
Sokolov-Ternov effect has been further emphasized and discussed.
}
while the Bogolubov transformation is trivial and the vacuum
associated to circular motion is still the Minkowski vacuum. In fact, in \cite{Letaw:1980ik}
the same authors quantized a scalar field for all possible choices of stationary coordinate systems
in Minkowski space and found that there are only two possible vacuum states: the Minkowski
vacuum (for coordinate systems without an acceleration horizon) and the Fulling vacuum (for systems with an acceleration horizon).
That the existence of a horizon is irrelevant for the Unruh-DeWitt detector to click was also nicely
shown in \cite{Korsbakken:2004bv}: by considering simultaneously rotation and acceleration and tuning
$a$ (linear acceleration) and $\omega$ (angular velocity), it is possible to move continuously between situations with and without horizon while
the outcome of the Unruh-DeWitt detector does not suffer dramatic changes.

In \cite{Davies:1996ks}, Davies, Dray and Manogue focused in the rotating case and argued that 
the results from Bogolubov transformations and the Unruh-DeWitt detector should be reconciled
whenever one is in a situation in which the rotating Killing vector never becomes space-like,
{\it i.e.} in cases in which no point of space-time co-rotating with the detector exceeds the
speed of light. In order to support this statement, they studied the behavior of an Unruh-DeWitt detector undergoing circular motion in 
a bounded space ({\it e.g.} an infinite cylinder or the Einstein universe). The results are very interesting. They found that if a detector is moving with an angular velocity
$\omega$ in a bounded space of radius $R$, then it will remain inert as long as $\omega R < a_0$, where $a_0$ is some constant bigger than one\footnote{The precise value of $a_0$ depends on the geometry of the space.}. On the contrary, whenever $\omega R > a_0$ the detector will ``click'' and moreover the spectrum of radiation will not be thermal. 
\vskip.1cm

The problem of ``detectors clicking'' ({\it i.e.} non-trivial interaction with the quantum vacuum) still has several aspects that deserve further study. 
Moreover, most of the results presented in the lines above where obtained for weakly coupled field theories and some of them rely deeply on 
techniques that are only valid in this regime. 
Going to the strongly coupled case could be a serious challenge but as we will show in this work, the $AdS$/CFT correspondence \cite{Maldacena:1997re} 
allows us to approach this problem in a very simple manner. 

We will study accelerated detectors by means of the $AdS$/CFT correspondence and try to provide some complementary insight
to the question: {\it what does an observer in non-geodesic motion detect?}
Along the way we will also find some interesting results related to the dynamics of an infinitely massive quark living in a $S^3$.
We define the detector probing the quantum vacuum of the CFT as an external quark in $\mathcal{N}=4$ SYM. The dual picture corresponds to an open string that extends radially from the boundary to the horizon of $AdS$ \cite{Maldacena:1998im}. To be more precise, the quark corresponds to the tip of a string, whose body codifies the profile of the non-Abelian fields sourced by the quark. 

Apart from the theoretical interest, an accelerated probe in a strongly coupled system may be of phenomenological 
interest, since it has been recently suggested \cite{Tuchin:2010vs} that acceleration of quarks
induced by the huge magnetic field produced in heavy ion collisions can have measurable consequences. 

The main goal of this work is to put forward the idea that the holographic 
signature of a non-trivial response of a probe to the quantum vacuum  is the presence
of non-trivial Hawking radiation propagating in the string worldsheet, namely the existence of a
worldsheet horizon. In a sense, this is the analog of a non-vanishing
Unruh-de Witt detection rate. However, the systems under discussion have physical differences 
with respect to those
in which the Unruh-de Witt computation is applicable (in particular, we remain at strong coupling,
as already pointed out)
and therefore this analogy should not be taken
too far away.

In the context of the $AdS$/CFT correspondence, an important observation    
regarding the role of worldsheet horizons was made
 in \cite{deBoer:2008gu, sonteaney}\footnote{An early discussion of
the importance of horizons for the physics of probes was presented in \cite{Kiritsis:1999tx}.}. The physics of a finite mass quark sitting in a thermal bath (in a deconfined phase) was studied by analysing a straight string
in an $AdS$-Schwarzchild background. The string worldsheet stretches from the black hole horizon up to a UV cut-off near the boundary.
Semiclassical Hawking radiation emanating from the horizon populates the various modes of oscillation of the string, 
thereby making the endpoint/quark jitter. In this case, the stochastic motion of the quark is correctly described by a generalized Langevin equation. It is this kind of random processes what will make the detector click. 

If the string is in non-trivial motion, the
worldsheet horizon does not have to coincide with the bulk horizon.
As emphasized in \cite{Giecold:2009cg, CasalderreySolana:2009rm}
 (and also shown in \cite{Gursoy:2010aa} for a non-conformal theory where the notion
 of strings as moving thermometers was also introduced), the random processes affecting the physics
of the quark in the dual theory are controlled, in this case, by the worldsheet horizon.
Crucially,  non-geodesic motion of the string endpoint lying at the $AdS$
boundary, can produce a worldsheet horizon even when the bulk geometry does not have a 
horizon \cite{Chernicoff:2008sa}.
For the particular case
when the endpoint moves with uniform proper acceleration, 
this  issue has been discussed in \cite{Dominguez:2008vd, Xiao:2008nr, Paredes:2008cr}
--- see also \cite{Hirayama:2010xi,Ghoroku:2010sp,BatoniAbdalla:2007zv} for related work.
The corresponding worldsheet horizon has a temperature that is precisely given by (\ref{unruh}), 
and there is also an 
associated brownian motion \cite{Caceres:2010rm}. 

In this paper, we will generalize this observation to arbitrary motions of the string endpoint.
Moreover,
we will consider the field theory living in both $\RR^{1,3}$ and $\RR \times S^3$ and 
emphasize the physical differences
that stem from having a finite space. 

Knowing the string embedding
is an essential ingredient of this analysis.
We begin in section \ref{Mikhailov} by reviewing a remarkable paper by Mikhailov \cite{Mikhailov:2003er}, who constructed an analytic embedding for a string in $AdS$
dual to an infinitely-massive quark in $\mathcal{N}=4$ SYM that follows an \emph{arbitrary} timelike trajectory. 
The solution is valid for both global $AdS$ and the Poincar\'e patch but it was written
explicitly only for the latter. To make contact with the results of \cite{Davies:1996ks}, we 
will work in global $AdS$ where the dual boundary theory lives in a $S^3$. In section \ref{globalAdS} we
 work out the solution for the string embedding and calculate the on-shell energy of the string. 
As in the Minkowski case \cite{Chernicoff:2008sa}, the total energy of the string splits into two terms: $E_q$ given by (\ref{energyq}) which appears as total derivative and is associated to the
intrinsic energy of the quark (and anti-quark) at time $t$, and $\mathcal{P}_q$ written in (\ref{pq}) that represents the energy \emph{lost} by the quark
over all times prior to $t$. Equation (\ref{pq}) is a Li\'enard like formula for a quark moving in a stongly coupled gauge theory living in a $S^3$. We close this section with a discussion regarding the concept of radiation in a strongly coupled theory defined on a bounded curved space. 

In section  \ref{sec: general}, armed with the analytic solution found in section \ref{Mikhailov}, we carefully scrutinize the structure of the induced worldsheet metric. We show that for an arbitrary motion of the string endpoint and as long as the squared
proper acceleration $A_6^2$ defined in (\ref{A62}) is a positive
constant, there is a worldsheet horizon with a temperature given by
\be
T_{ws}=\frac{A_6}{2\pi}\,.
\ee
We would like to emphasize that
it is remarkable that this Unruh-like formula holds for general motions and not just for
 the case of uniform acceleration. It is also important to keep in mind that this 
 worldsheet temperature is not necessarily identified with a temperature in the gauge
  theory; we will briefly comment on this matter in appendix \ref{sec:GEMS}.
When $A_6^2<0$ there is not a worldsheet horizon and it does not make sense to define $T_{ws}$.
We will distinguish between cases with and without horizon, but computing the 
precise detector excitation
rates lies beyond the scope of this work.

Circular motion will be the subject of section
\ref{sec: circular}. The solution for a rotating string in the Poincar\'e patch
has been recently discussed in
\cite{Athanasiou:2010pv}. Indeed, there is a worldsheet horizon, meaning that the quark
will measure non-trivial fluctuations of the vacuum. On the other hand, the coordinate change
adapted to the rotating quark does not produce an acceleration horizon (see appendix \ref{app:coordchange}). Following
the idea of Davies {\it et al.} \cite{Davies:1996ks}, the apparent paradox should fade away
when the theory lives in a bounded space, such that no co-rotating observer would exceed the
speed of light. In order to bear out this observation, 
we restrict the solution (\ref{rhoglobalsolution}) and consider the particular case of a string rotating in global $AdS$. 
The string embedding in given by (\ref{eqR}) and (\ref{eqfi}). The dual picture is a quark rotating parallel to an
equator of the $S^3$. We discuss a family of solutions depending on the latitude $R_0$ and 
the angular velocity $\omega$. There is indeed a transition between solutions with and without worldsheet
horizons. Therefore, our results are in qualitative agreement with those of \cite{Davies:1996ks} 
even though we are working
in a strongly coupled, non-abelian theory. 

Throughout the paper, we have done all the calculations considering an infinitely heavy quark which might seem 
awkward since stochastic motion is suppressed in the large mass limit. However,
we are only interested in qualitative results, and, in particular, in finding the position of the worldsheet horizon. From \cite{Chernicoff:2009re}, we know that Mikhailov's solution can be written for a finite mass quark as well
and, qualitatively, our conclusions remain the same. We will briefly come back to this subtlety in section 
\ref{sec: general}.

We believe that the fact that we can provide some insight into these old questions in quantum field
theory from a different perspective constitutes yet another illustration
of the usefulness of the $AdS$/CFT correspondence.

\section{Strings moving in $AdS_5$}
\label{Mikhailov}
\setcounter{equation}{0}

In this section, we will review and extend the work of \cite{Mikhailov:2003er}, which
will be an essential tool in the subsequent discussion.
The author of this paper managed to find explicitly the open string worldsheet embeddings solving
the Nambu-Goto equation inside anti-de Sitter space, for an arbitrary time-like trajectory of the
string endpoint attached to the boundary; namely for an arbitrary time-like trajectory of the quark
in the dual theory. It can be interpreted as a non-linear wave propagating in the worldsheet for
an arbitrary perturbation.

It is important to remark that it is not the most general solution: the boundary
condition is given such that the wave flows away from the quark and never comes back. In particular,
in global $AdS$, this means that an anti-quark lying at the antipode of the quark absorbs completely
the perturbation.

Consider $AdS_5$ space as a hyperboloid embedded in $\RR^{2,4}$, namely
\be
X^M X_M = (X^{-1})^2 + (X^{0})^2 - (X^{1})^2 - (X^{2})^2 - (X^{3})^2 - (X^{4})^2 = 1
\label{hyperboloid}
\ee
where the $M$ indices run over the six dimensional target space $M=-1,0,\dots, 4$
and we define $\eta_{MN} = \{++----\}$ with mostly minus notation.
We have taken the $AdS$ radius to one, but we will eventually reinsert it when 
computing some quantities in the dual field theory, taking into account that
$R_{Ads}^2/\alpha' = \sqrt{\lambda}$, where $\lambda$ in the 't Hooft coupling.

For any vector depending on one parameter
$l^M(\tau)$ and satisfying
\be
l^M l_M = 0 \,\,,\qquad\qquad (\partial_\tau l^M) (\partial_\tau l_M)=1
\label{ldef}
\ee
there is a surface --- worldsheet --- which is parameterized by $\tau,\sigma$ which is 
extremal within $AdS$ (namely, the string solves the equations from the Nambu-Goto action) and can be written as
\be
X^M (\tau,\sigma) = \pm \partial_\tau l^M(\tau) + \sigma \, l^M(\tau)\,\,.
\label{vofl}
\ee
Notice
that using (\ref{ldef}), it is immediate to show that $X^M X_M = 1$, namely
this surface indeed lies within $AdS_5$.
In the gauge theory language, the $\pm$ sign of (\ref{vofl}) corresponds to the choice between a purely outgoing or purely ingoing boundary condition for the waves in the gluonic field at the location of the source. We will focus solely on the solution with the plus sign, which is the one that captures the physics of present interest, with influences propagating outward from the quark to the anti-quark.

The induced metric on the worldsheet was also given in \cite{Mikhailov:2003er} and takes the simple form
\be
ds_{ws}^2 = ( \partial^2_\tau l_M \partial^2_\tau l^M + \sigma^2 )d\tau^2 - 2 d\tau d\sigma.
\label{dsws2}
\ee

\subsection{Strings in the Poincar\'e patch}

In order to proceed further, let us give an explicit realization of the 
$X^M$ satisfying (\ref{hyperboloid}). If we write
\be
X^{-1} = \frac{1}{2z} (1-(x_\mu)^2 +z^{2}) \,\,,\qquad
X^\mu = z^{-1}\,x^\mu\,\,,\qquad
X^4 = \frac{1}{2z} (1+ (x_\mu)^2 - z^{2}) \,\,
\label{PoincareGEMS}
\ee
we are in the Poincar\'e patch. The dual CFT lives in $\RR^{1,3}$.
The greek index runs over $\mu=0,1,2,3$ 
and we will contract it with a 4-dimensional mostly minus Minkowski metric
$\eta_{\mu\nu} = \{+---\}$.  
The Poincar\'e metric on $AdS$ follows
\be\label{poincaremetric}
ds_5^2 = \eta_{MN} dX^M dX^N = \frac{1}{z^2}\left(
\eta_{\mu\nu} dx^\mu dx^\nu - dz^2
\right)
\ee
with the boundary at $z=0$.
Let us write down the vector $l^M(\tau)$ in terms of
a set of four arbitrary functions $\bar x^\mu (\tau)$,
to be  identified with the
trajectory of the string endpoint attached to the boundary
\be
l^{-1}= \frac12\left(1- (\bar x_\mu(\tau))^2 \right)\,\,,\qquad
l^\mu = \bar x^\mu(\tau)\,\,,\qquad
l^4 = \frac12\left(1+ (\bar x_\mu(\tau))^2 \right)\,\,.
\label{lPoincare}
\ee
Regarding (\ref{ldef}), we need
\be
\partial_\tau {\bar x}^\mu\, \partial_\tau {\bar x}_\mu = 1 \,\,,
\label{propertau}
\ee
such that $\tau$ is the proper time of the endpoint/quark trajectory. 
Notice that $\sigma$ is fixed and equal to $\sigma = z^{-1}$, then equations (\ref{vofl}) are reduced to:
\be
x^\mu(\tau,z) = z\, \partial_\tau {\bar x}^\mu(\tau) + \bar x^\mu (\tau)\,\,.
\label{PoincareSolution}
\ee

Equation (\ref{PoincareSolution}) displays the
string worldsheet as a ruled surface in $AdS_5$, spanned by the straight lines at constant
$\tau$. 

To be more concrete, from the
$\mu=0$ component of (\ref{PoincareSolution}), parametrizing the quark worldline by $\bar{x}^0(\tau)$
instead of $\tau$, and using $d\tau=\sqrt{1-\vec{v}^{2}}d\bar{x}^0$, where $\vec{v}\equiv
d\vec{\bar{x}}/d\bar{x}^0$, this amounts to
\be\label{tret}
t=z{1\over\sqrt{1-\vec{v}^{\,2}}}+ t_{\text{ret}} ~,
\ee
where we have introduced
$t\equiv x_0$, $t_{ret}=\bar x_0$ and
 the endpoint velocity $\vec{v}$ is meant to be evaluated at $\tret$. In these same
terms, the spatial components of (\ref{PoincareSolution}) can be formulated as
\be\label{xmikh}
\vec{x}(t,z)=z{\vec{v}\over\sqrt{1-\vec{v}^{\,2}}}+\vec{x}(t_{\text{ret}})=(t-t_{\text{ret}})\vec{v}+\vec{x}(t_{\text{ret}})~.
\ee
As explained in
\cite{Chernicoff:2008sa}, $\tret$ should be understood as a retarded time.

\subsubsection*{An example: rotating string in the Poincar\'e patch}
\label{example}

Later on this letter we will discuss the physics of a quark rotating in Minkowski space and compare it to the case of a quark rotating in a $S^3$ 
(dual to a string in the Poincar\'e patch and global $AdS$ respectively) which, as we have mentioned in the introduction, is our major concern.
With this in mind, we will now obtain the solution found in \cite{Athanasiou:2010pv} in which the
string endpoint is rotating in a circle of radius $R_0$ at constant
angular velocity $\omega$, and show that it is a particular case of the formalism presented in the previous section. 
Let us write
\be
\bar x^1= R_0 \cos(\omega\, \tret(\tau)) \,\,,\qquad\qquad
\bar x^2 = R_0 \sin(\omega\, \tret(\tau)) \,\,,\qquad\qquad
\bar x^3=0;
\ee
and from (\ref{propertau}), we read the relativistic factor
\be
\partial_\tau \tret = \frac{1}{\sqrt{1-R_0^2\omega^2}} \equiv \gamma
\ee
and therefore $\tret=\gamma\,\tau$.
We now compute the string worldsheet by inserting these boundary conditions into
(\ref{PoincareSolution})
\bear
t(\tret, z)=z \,\gamma + \tret\,\,,\qquad
x^1(\tret,z)&=&-z\,\gamma\,R_0\omega\sin(\omega\tret)+R_0\cos(\omega\tret)\,\,,\rc
x^3(\tret,z)=0\,\,,\qquad\quad x^2(\tret,z)&=& z\,\gamma\,R_0\omega\cos(\omega\tret)+R_0\sin(\omega\tret)\,\,.\rc
\label{tretliu}
\eear
If we define polar coordinates in the $x^1-x^2$ plane according to
$x^1 = R \,\cos\phi$, $x^2=R\,\sin\phi$, we can rewrite the above result as
\bear
R^2(t,z) &=& R_0^2 (1+z^2 \omega^2 \gamma^2)\,\,,\rc
\phi(t,z) &=& \omega (t-\gamma z)+ \arctan(\omega \,\gamma z)\,\,,
\label{Athana}
\eear
where we have written $\phi$ in terms of coordinate time using the first expression
of (\ref{tretliu}). The solution (\ref{Athana}) has been discussed in \cite{Athanasiou:2010pv}.

\subsection{Strings moving in global $AdS$}\label{globalAdS}

One can solve (\ref{hyperboloid}) in terms of a different set of coordinates
\be
X^{-1}=\cos \left(\frac{t}{b}\right)\,\sqrt{1+\left(\frac{r}{b}\right)^2}\,\,,\quad X^0=\sin \left(\frac{t}{b}\right)\,\sqrt{1+\left(\frac{r}{b}\right)^2
}\,\,,\quad
X^a = \frac{r}{b}\,n^a  \,\,,
\label{globalGEMS}
\ee
where $b$ is a constant and 
$a=1,\dots,4$ and the $n^a$ span the surface of a unit $S^3$,
namely
\be
\sum_{a=1}^4 (n^a)^2 = 1\,\,.
\label{constrainna}
\ee
The $n^a$ could be realized in terms of some particular set of angles parameterizing
the sphere but for the moment it is convenient to keep them general.

These are the global coordinates and when we consider $AdS$ written in this
fashion, the dual CFT lives in a three-sphere $\RR \times S^3$ 
(with the real line corresponding to global time).
The global $AdS$ metric (with unit $AdS$ radius) is
\be
ds_5^2 = \eta_{MN} dX^M dX^N = b^{-2}\left[
(1+(r/b)^2)\ dt^2 - \frac{dr^2}{1+(r/b)^2} - r^2\ d\Omega^2_3 \right]
\label{globalmetric}
\ee
with the boundary at $r=\infty$ and where $d\Omega^2_3=dn^a dn^a$ is the metric of the unit $S^3$. 
The constant $b$ is the radius of the $S^3$ in which the dual field theory lives
and we will also set it to 1 in the following in order to reduce cluttering of notation.
We will however reinsert it on dimensional grounds at several points for the interpretation
of results.

The next step is to find a parameterization of $l^M (\tau)$ which, as in the Poincar\'e case,
is written in terms of a set of functions  which can eventually
be identified with the motion of the endpoint at the boundary.
Let us write
\be
l^{-1}= \cos (\tret(\tau))\,\,,\qquad
l^0 = \sin (\tret(\tau))\,\,,\qquad
l^a = \bar n^a (\tau)\,\,.
\label{lglobal}
\ee
If we impose the constrains
\be
\sum_a (\bar n^a)^2 = 1\,\,,\qquad
\partial_\tau {t^2_{\text{ret}}} - \sum_a  (\partial_\tau {\bar n}^a(\tau))^2 = 1\,\,,
\label{globalttnn}
\ee
the conditions (\ref{ldef}) are automatically solved.
Defining
\be
v^a = \partial_{\tret} \bar n^a\,\,,\qquad
a^a = \partial_{\tret} v^a\,\,,
\label{vaaadefs}
\ee
the second equation of (\ref{globalttnn}) require
\be
\partial_\tau \tret= \gamma \,\,,\qquad
\gamma = \frac{1}{\sqrt{1-v^2}}\,,
\label{gammaglobal}
\ee
where $v^2 \equiv v^a v^a $ and we have introduced the relativistic factor $\gamma$ which for
a general motion will depend on $\tret$.
Notice that $v^a$, $a^a$ are defined as velocity and acceleration in the $\RR^4$ space in 
which the $S^3$ is embedded, and not as velocity and acceleration on the sphere itself\footnote{
Nevertheless, $v^2$ is the same in
$\RR^4$ and $S^3$.}.

Inserting (\ref{lglobal}) into
(\ref{vofl}) --- where again we take the plus sign ---, we find 
the worldsheet embedding
\bear
X^{-1}&=&
- \,\gamma\,\sin \tret + \sigma \,\,\cos \tret
\,\,,\rc
X^0 &= &
\gamma\,\cos \tret + \sigma \,\sin \tret
\,\,,\rc
X^a &=& \gamma\,v^a + \sigma \, \bar n^a\,\,.
\eear
We now relate this expression to the $t,\ r,\ n^a$ coordinates of global
$AdS$ through (\ref{globalGEMS}), with $b=1$. 
After some algebra one gets
\bear
t(\tret, \sigma) &=& \tret + \arctan \left(\frac{\gamma}{\sigma}\right)\,\,,\rc
r(\tret, \sigma) &=& \sqrt{\sigma^2 + \gamma^2 v^2}\,\,,\rc
n^a(\tret, \sigma) &=& \frac{1}{r}\left(
\gamma\,v^a + \sigma \, \bar n^a \right)\,\,.
\label{rhoglobalsolution}
\eear
We have written the embedding in terms of coordinates $\tret$, $\sigma$,
but it can be also understood in terms of the original worldsheet coordinates 
$\tau,\sigma$ by noticing the $\tret$ relation implicit in (\ref{gammaglobal}).
If we want both  endpoints of the string to reach
the boundary $r=\infty$, it is clear that 
the worldsheet coordinate has to range in $\sigma \in (-\infty,\infty)$.
For any given $\tau$,  the two endpoints lie at antipodal points
\be
n^a(\tau, +\infty)= - n^a(\tau, -\infty) = \bar n^a(\tau) \,.
\ee
where we have taken into account that 
$\lim_{\sigma \to \pm\infty} \frac{r}{\sigma}=
\pm  1$. Thus, the solution is representing an infinitely massive
 quark at a given point of the sphere and an
anti-quark at the antipode.

\subsubsection*{The energy of the string and a Li\'enard-like formula}
\label{lienardlike}

Another important result of \cite{Mikhailov:2003er} is the emergence of Li\'enard's
formula for the power radiated by an accelerated charge that is obtained by computing
the on-shell energy of the string worldsheet in the Poincar\'e patch. 
In this section, we generalise this computation to the case of global $AdS$, which
should be associated to the energy flowing away from the charge moving on a three-sphere\footnote{It is important to be careful with the concept of radiation and energy loss in a bounded space. We will comment on this further in the text.}.

Concretely, we want to compute the on-shell energy density of the string
{\it at fixed global coordinate time} $t$.
With that aim, we parameterize the
worldsheet as $n^a (t,\sigma)$, $r(t,\sigma)$, such that the Nambu-Goto action is
\be
S =-\frac{1}{2\pi\alpha'} \int \sqrt{{\textrm {-det}}( P[g])}d\tau d\sigma
 = \frac{\sqrt{\lambda}}{2\pi} \int {\cal L}\, dt\,d\sigma
\ee
where we have reinserted the $AdS$ radius ('t Hooft coupling of the dual theory)
and introduced the lagrangian density\footnote{The expression (\ref{lagrglob})
cannot be directly used to extract equations of motion for the fields
$n^a (t,\sigma)$, $r(t,\sigma)$, since the $n^a$ are subject to a constraint
(\ref{constrainna}). Nevertheless, once we know the solution,
(\ref{lagrglob}) can be used to compute the on-shell lagrangian and hamiltonian densities.}
\bear
{\cal L}=-\Bigg(
(\partial_\sigma r)^2 - \frac{r^2}{1+r^2} (\partial_t n^a)^2(\partial_\sigma r)^2 + 
r^2(1+r^2) (\partial_\sigma n^a)^2-r^4 (\partial_t n^a)^2(\partial_\sigma n^b)^2+
\rc
-\frac{r^2}{1+r^2}(\partial_t r)^2(\partial_\sigma n^a)^2+
r^4 ( \partial_t n^a \partial_\sigma n^a)^2+\frac{2r^2}{1+r^2}
( \partial_t n^a \partial_\sigma n^a) ( \partial_t r)( \partial_\sigma r)
\Bigg)^{\frac12}\ \ \,
\label{lagrglob}
\eear
The energy is an integral of the hamiltonian density
\bear
 E(t) 
&=&\frac{\sqrt{\lambda}}{2\pi} \int_{-\infty}^{\infty} {\cal H} d\sigma =
\frac{\sqrt{\lambda}}{2\pi} \int_{-\infty}^{\infty} \left((\partial_t n^a)\frac{\partial {\cal L}}{\partial (\partial_t n^a)}
+(\partial_t r)\frac{\partial {\cal L}}{\partial (\partial_t r)} - {\cal L}
\right) d\sigma \rc
&=&-\frac{\sqrt{\lambda}}{2\pi} \int_{-\infty}^{\infty}  \frac{(\partial_\sigma r)^2+ 
r^2(1+r^2) (\partial_\sigma n^a)^2}{{\cal L}} d\sigma
\label{Eglobal}
\eear
We now evaluate this expression by inserting the solution (\ref{rhoglobalsolution}).
This is a rather lengthy but straightforward computation, in which the final result takes
a remarkably simple form. Before discussing it, let us briefly comment on some intermediate
steps.
A subtlety is that in
(\ref{Eglobal}),  the partial derivative $\partial_\sigma$ has to be taken at
fixed $t$ whereas $\partial_t$ has to be taken at fixed $\sigma$. On the other hand the
quantities in
(\ref{rhoglobalsolution}) are written as a function of $\sigma$, $\tret$. Therefore, 
they should be interpreted
as functions of $\sigma$, $\tret(t,\sigma)$ where 
the $\tret(t,\sigma)$ relation is implicitly given by the
first expression of (\ref{rhoglobalsolution}). 

Another important point is that it is better to reexpress the energy as an integral over 
$\tret$. We use the first equation in (\ref{rhoglobalsolution}) 
at fixed $t$ to change from $\sigma$ to $\tret$. 

Taking all this into account, (\ref{Eglobal}) gives
\be
E(t) = \frac{\sqrt{\lambda}}{2\pi}\int_{t-\pi\,b}^t 
d\tret {\cal P}_q + E_q(\vec{v}(t))
\label{lienardglobal}
\ee
where the first term is the rate of energy flowing down the string from one endpoint (quark) to the other (anti-quark)
and is given by
\be\label{pq}
{\cal P}_q = \frac{a^2(1- v^2) + (\vec v \cdot \vec a)^2 - v^2 (1-v^2)/b^2}{(1-v^2)^3}.
\ee
In this expression
we have reinserted the constant $b$ and defined
\be
v^2 = v^a v^a\,\,,\qquad
(\vec v \cdot \vec a) = v^a a^a\,\,,\qquad
a^2 = a^a a^a\,\,.
\label{v2a2defs}
\ee
The integral over
$\tret$ encodes the accumulated energy \emph{lost} by the quark
over all times prior to $t$. This term is a generalization of Li\'enard's formula for a quark living in a strongly coupled theory defined in the $S^3$. 
As $b \to \infty$, the result of \cite{Mikhailov:2003er} --- Li\'enard's formula --- is recovered.
Clearly (\ref{pq}) is not positive definite which might seem strange
and we will later comment on this matter.

The second term in (\ref{lienardglobal}) comes from a total derivative and following the discussion of \cite{Chernicoff:2008sa}, it should
be associated to the dispersion relation of an infinitely massive quark 
(and  infinitely massive anti-quark). The result is
\be\label{energyq}
E_q=\frac{\sqrt{\lambda}}{2\pi} \gamma\, \sigma \Big{|}^{\sigma=\infty}_{\sigma=-\infty}\,\,,
\ee
which is the
intrinsic energy of the quark (and anti-quark) at time $t$.
Notice that as in the Minkowski case, the split of the total energy of the string
admits a clear and pleasant physical interpretation.

In the Poincar\'e patch, the dynamics of
 a finite mass quark --- such that $\sigma$ ranges in a finite interval ---
  was studied in \cite{Chernicoff:2009re, Chernicoff:2009ff}. The authors found a nonlinear generalization of the Lorentz-Dirac equation that correctly incorporates the effects of radiation damping. It would be interesting to generalise this computation to global $AdS$, but that lies beyond the scope of the present work.

For a quark rotating in Minkowski space
(namely a string rotating in the Poincar\'e patch), the emitted radiation was studied in 
great detail in \cite{Athanasiou:2010pv} and it was found that the radiated power is very similar to that of synchrotron radiation produced by an electron in 
circular motion in classical electrodynamics. 
When the theory lives on the $S^3$, the situation is different. First of all,
since one cannot go to asymptotic infinity, it is not clear how to define radiation. 
However, the boundary conditions implicit in Mikhailov's solution (\ref{vofl}) imply that all the energy
released by the quark is absorbed by the antipodal antiquark. As argued in 
\cite{Sin:2004yx}, this is the most similar
 situation to what one can call radiation on the sphere. 
Loosely speaking, one
could think of the situation as adding a point to $R^3$ at infinity, such that it becomes topologically $S^3$.
This ``point at infinity", where the antiquark 
is, does not allow radiation to bounce back. Therefore, whatever
gluonic field is radiated, it only lives for a finite time, before it reaches the antiquark and is absorbed, a time
$\pi\,b$ later, as encoded in the integration limits of (\ref{lienardglobal})\footnote{
This approximate notion of radiation loses its validity when the size of the emitted pulse 
is comparable to the travelled distance, namely the size of the sphere. We will further discuss
this issue in section  \ref{sec: circular} in a particular case.}. The time of flight of 
a signal propagating through the worldsheet between the two endpoints is the same
as that of a signal propagating at the speed of light in the dual theory from one point of the sphere to the antipode,
which is in fact $\pi\,b$.  We remark that, for general quark-antiquark 
string extended in the bulk, the time of flight via the worldsheet and via the boundary
do not match \cite{Bak:1999iq, Callan:1999ki}. 
Consequently, keeping in mind the aforementioned words of caution,
 it is natural to interpret the first term in (\ref{lienardglobal}) as the radiation produced
by an electric point-like charge living on a $S^3$.
 
In \cite{Villarroel:1974hm}, it was argued that
the generalization of Li\'enard's formula to curved spaces should simply be given --- apart from an overall
multiplicative constant --- by the square of the proper acceleration
\be
\frac{dE}{dt} = A_4^2 = - \frac{DU^\alpha}{d\tau}\frac{DU_\alpha}{d\tau}
\label{Villarroel}
\ee
where $\tau$ is the proper time on the particle worldline, $U^\alpha$ is the four-velocity and 
$D$ stands for the covariant derivative computed using Christoffel symbols
\be
DU^\alpha = \frac{dU^\alpha}{d\tau} + \Gamma_{\beta\gamma}^\alpha U^\beta U^\gamma\,\,.
\ee
We have introduced the notation $A_4^2$ in order to make clear that this is a
proper acceleration
computed with the 4d metric, the index $\alpha$
takes values $0,\dots,3$ running over the coordinates of $ds_4^2 = dt^2 - b^2 d\Omega_3^2$.
By inserting this metric into (\ref{Villarroel}) and transforming to 
the notation above (namely $v$, $a$ correspond to velocity and acceleration as computed
in the $R^4$ in which the $S^3$ is embedded), a straightforward computation shows
\be
A_4^2 = \frac{a^2(1- v^2) + (\vec v \cdot \vec a)^2 - v^4 (1-v^2)/b^2}{(1-v^2)^3}\,\,.
\label{PVilla}
\ee
The expression (\ref{PVilla}) is similar to what we got in (\ref{pq}), but not the
same, the difference is $v^2 \gamma^2 b^{-2}$.
Let us briefly comment on this mismatch, for which strong coupling and the finite size of the sphere
seem to play an important role. First, the result of \cite{Villarroel:1974hm} in principle cannot be
used for the $S^3$ since its computation is for the energy that escapes to infinity, something that 
cannot happen in the compact $S^3$. And second, notice 
that (\ref{PVilla}) is positive definite whereas $\mathcal{P}_q$ given in
(\ref{pq}) is not. When ${\cal P}_q < 0$, there is an energy flow up the string from the
anti-quark to the quark.
This seems to reflect the fact that, at least in some regime\footnote{In section \ref{sec: circular} we will study the particular case of a rotating quark and see this explicitly.}, 
the quark and the anti-quark cannot be considered as independent objects.

\section{Worldsheet horizons and clicking detectors}
\label{sec: general}
\setcounter{equation}{0}

In the introduction, we have reviewed that a worldsheet horizon induces some rate of random motion on the
dual quark. However, when one thinks of Unruh-De Witt detectors, one typically thinks that what is
getting excited is some internal degree of freedom, such as for instance swapping the spin of an 
electron. In the $AdS$/CFT framework, there is a kind of internal degrees of freedom which one can
think of, in some sense, as the analog the electron spin. That is, the worldsheet Hawking radiation
will not only cause the endpoint/quark to fluctuate around its classical trajectory along the boundary of $AdS_5$, but also along the internal space,
which would be $S^5$ in the benchmark case of $AdS_5 \times S^5$. More precisely, one could think of a
finite mass quark as a string ending on a D7 brane which wrap an $S^3$ inside the $S^5$.
Following the previous arguments, the string 
endpoint associated to the quark will jitter along the $S^3$ if and only if there is a worldsheet horizon:
the internal degrees of freedom of the quark will be excited 
--- the detector will click with a non-trivial rate ---
only in that case. 
Thus, there will be stochastic motion in both space-time and the internal sphere.

At this point, it is worth commenting on the implications of considering an
infinitely massive quark. Clearly, fluctuations around the 
classical quark
trajectory are suppressed in this limit. Nevertheless, if one deals with a finite mass quark along the lines of 
\cite{Chernicoff:2009re}, the position of the horizon would remain unchanged, provided the $\bar x_\mu(\tau)$
(or ($\bar n_a(\tau)$) are reinterpreted accordingly. 
Moreover, one should keep in mind that
a finitely massive quark has a ``gluonic cloud" around it and cannot be reinterpreted as a point-like object
(as opposed to the usual Unruh-de Witt case). A more refined analysis of these features lies beyond the scope
of the present work.
On the other hand,
it is important to mention that the internal degrees of freedom will get excited even in the case of an 
infinitely massive quark.

In this regard, a somewhat similar situation was studied in \cite{Das:2010yw} where the 
authors considered a D1 fixed in the gauge theory directions but rotating with constant
angular velocity in one of the maximal circles of the $S^5$. In the case of the Poincar\'e patch there is indeed a worldsheet horizon and therefore an internal degree 
of freedom is being excited. They studied the same situation in global $AdS$ but only geodesic motion was considered, then there was not a worldsheet horizon and according
to our proposal the interaction with the quantum vacuum is trivial.
We just notice here that this stochastic motion on the internal directions must
exist, but we will not attempt here to explicitly compute its form.

As we have seen in the previous section, we have the embedding functions of the string for an arbitrary time-like trajectory of the string endpoint. We now want to use this result
and compute the induced metric for a rather general situation.

The induced metric in the worldsheet for an arbitrary trajectory of the enpoint $l^M(\tau)$ was
written in (\ref{dsws2}). We remind the reader that this is valid both for the Poincar\'e patch
and for global $AdS$, we will later comment separately on the two cases. 

Let us rename the quantity appearing in (\ref{dsws2}) as
\be
A_6^2 \equiv - \eta_{MN} \frac{d^2 l^M}{d\tau^2}\frac{d^2 l^N}{d\tau^2}
\ee
where the $A_6$ notation is introduced 
to make clear that this is a proper acceleration in the 6d space
defined by the $l$'s.
Using the notation introduced in section \ref{globalAdS}, one can easily check that
\be
A_6^2 = \frac{a^2 (1-v^2) + (\vec v \cdot \vec a)^2- (1-v^2)/b^2}{(1-v^2)^3}\,\,.
\label{A62}
\ee

This quantity  in general depends on $\tau$ but 
let us consider the case in which it is a constant
such that the worldsheet metric is stationary\footnote{
For non-constant $A_6^2$, one should study Hawking radiation with non-stationary geometries.
This would allow to study from $AdS$/CFT the effect of vacuum quantum fluctuations of non-stationary
motions as for instance that recently considered in
\cite{Kothawala:2009aj} from the Unruh-De Witt perspective.}. 
Performing the coordinate change
\be
\tau = \tilde \tau - \frac{1}{A_6} \textrm{arctanh} \left(\frac{\sigma}{A_6}\right)
\ee
the worldsheet metric takes the form
\be
ds_{ws}^2 = (\sigma^2 - A_6^2) d\tilde \tau^2 - \frac{d\sigma^2}{\sigma^2 - A_6^2}
\label{dsws2again}
\ee
which describes (a piece of) $AdS_2$. If $A_6^2>0$, there exists a horizon with temperature
\be
T_{ws} =  \frac{A_6}{2\pi}\,\,.
\label{unruhgen}
\ee
Notice that it is possible to have $A_6^2<0$ since $A_6$ corresponds to an acceleration in 
the 6d space-time spanned by the $l^M$ which has two time directions.

\subsubsection*{Penrose diagrams}
\label{penrose}

Let us be more precise about the causal structure on the worldsheet. It is useful 
to change to Kruskal-like coordinates in order to draw the associated
Penrose diagrams. We will first consider the string living in global $AdS$ and comment
at the end on what happens in the Poincar\'e patch.
We have to study separately the cases with $A_6^2 > 0$ and $A_6^2<0$.

\begin{figure}[htb]
\setlength{\unitlength}{1cm}
\includegraphics[width=6cm]{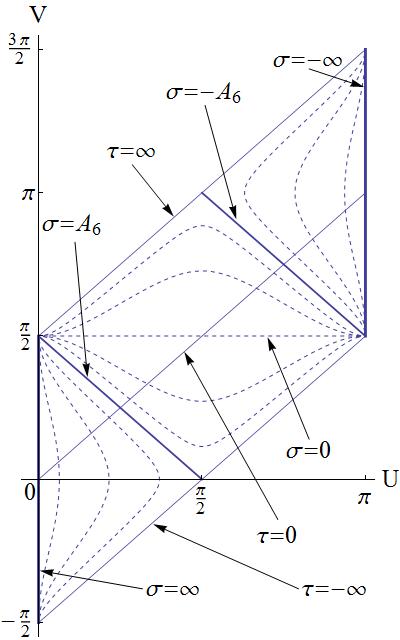} 
\includegraphics[width=6cm]{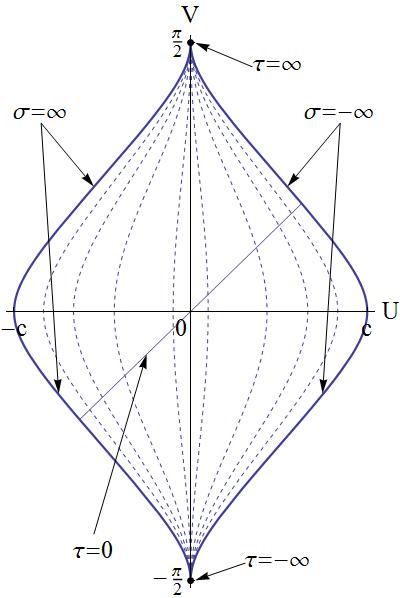}
	\caption{Penrose diagrams of the worldsheet
	for $A_6^2>0$ (left) and $A_6^2<0$ (right). Dashed lines are lines with constant
	$\sigma$. The two endpoints of the worldsheet correspond to the quark (at $\sigma=\infty$) and the antiquark
	(at $\sigma=-\infty$), which we depict with thick lines. In the second plot, they intersect
	the horizontal axis at $U=\pm c=\pm \arctan (\pi/2)$.
	Null trajectories going to the right correspond to constant $\tau$ in both plots. }
	\label{fig:penrose}
\end{figure}

When $A_6^2 > 0$, the appropriate coordinate change has to be defined piecewise:
\bear
\sigma > A_6 &\Rightarrow& 
\begin{cases}
V= \frac12 \left( \arctan (\tau +\frac{1}{A_6}\log \frac{\sigma + A_6}{\sigma - A_6} ) + \arctan \tau
\right) \cr
U =\frac12 \left( \arctan (\tau +\frac{1}{A_6}\log \frac{\sigma + A_6}{\sigma - A_6} ) - \arctan \tau
\right)
\end{cases} \rc
 |\sigma| < A_6 &\Rightarrow& 
\begin{cases}
V= \frac12 \left(\pi -\arctan (\tau +\frac{2}{A_6}\textrm{arctanh}\,\frac{\sigma}{A_6} ) + \arctan \tau
\right) \cr
U =\frac12 \left(\pi - \arctan (\tau +\frac{2}{A_6}\textrm{arctanh}\,\frac{\sigma}{A_6} ) - \arctan \tau
\right)
\end{cases}\rc
\sigma < - A_6 &\Rightarrow& 
\begin{cases}
V= \frac12 \left(2\pi + \arctan (\tau +\frac{1}{A_6}\log \frac{\sigma + A_6}{\sigma - A_6} ) + \arctan \tau
\right) \cr
U =\frac12 \left(2\pi + \arctan (\tau +\frac{1}{A_6}\log \frac{\sigma + A_6}{\sigma - A_6} ) - \arctan \tau
\right)
\end{cases} \nonumber
\eear
where we define the $\arctan$ to always take values between $-\pi/2$ and $\pi/2$.
In the new coordinates, the metric (\ref{dsws2}), for constant $A_6^2=-\partial^2_\tau l^M \partial^2_\tau l_M$
is rewritten as
\be
ds_{ws}^2 = \frac{|\sigma^2 - A_6^2|}{\cos^2(V+U) \cos^2(V-U)} \left(dV^2 -dU^2 \right)
\label{kruskalds}
\ee
which, as we wanted, is conformal to a flat metric spanned by $V$ and $U$, both bounded coordinates.
The associated Penrose diagram is depicted in figure \ref{fig:penrose} (left), in which
the ranges of $U,V$ associated to the full possible range of $\tau$ and $\sigma$ encode the
physics. The presence of causal
horizons is clear from the plot: a signal from the quark can reach the antiquark but the converse
is impossible. It is worth pointing out that such a signal from the quark to the antiquark satisfying
$dU=dV$ arrives in finite global time $t$
 (it takes a time $\Delta t = \pi$). On the other hand, trajectories
$dU=-dV$ would take infinite global time $t$ to reach the $\tau=\infty$ line, or to come
out from $\tau=-\infty$. A signal emitted from the anti-quark 
 would propagate towards larger $\sigma$'s but would spend infinite global time to reach
 $\sigma = -A_6$.

Let us now consider the case with $A_6^2 < 0$. With the coordinate change
\bear
V= \frac12 \left(\arctan (\tau -\frac{2}{\sqrt{-A_6^2}}\arctan\,\frac{\sigma}{\sqrt{-A_6^2}} ) + \arctan \tau
\right) \cr
U =\frac12 \left( \arctan (\tau -\frac{2}{\sqrt{-A_6^2}}\arctan\,\frac{\sigma}{\sqrt{-A_6^2}} ) - \arctan \tau
\right)
\eear
the metric can again be written as (\ref{kruskalds}). The Penrose diagram for this case is also shown in figure \ref{fig:penrose} (right). There are no causal
horizons and signals can go from the quark to the antiquark and vice-versa.
This is just the Penrose diagram of $AdS_2$.

If we consider the string living in the Poincar\'e patch, only $A_6^2>0$ is possible. Moreover, the 
string is just defined for $\sigma \geq 0$ (there is no antiquark). Thus, one just has to keep the
lower half of the plot on the left of figure \ref{fig:penrose}. Of course, there is still a horizon
at $\sigma=A_6$, no signal from $\sigma < A_6$ can reach the quark at $\sigma= \infty$.

\subsubsection*{Discussion}

We have already mentioned that when the bulk is the Poincar\'e patch of $AdS$, the dual theory lives in a flat 
$\RR^{1,3}$ space. One can think of it as the decompactification limit of $\RR\times S^3$
when the radius of the sphere diverges $b\to \infty$. Comparing (\ref{A62}) to (\ref{PVilla}),
we see that $A_6^2 = A_4^2$ in this limit and therefore (\ref{unruhgen}) tells us that the
worldsheet temperature has to be associated to the proper acceleration of the quark.
It seems that we are just recovering Unruh's formula (\ref{unruh}). However, the result is
very surprising because (\ref{unruhgen}) is valid for {\it any} stationary motion and not
just for uniform acceleration. Thus, in this sense, the situation at strong coupling looks simpler.
{
We will further comment on this fact in appendix \ref{sec:GEMS}, but a deeper understanding of this phenomenon
requires further work.

When the dual theory lives in $\RR\times S^3$, it is not true any more that the proper acceleration gives the
worldsheet temperature since $A_6 \neq A_4$. In fact, from 
(\ref{PVilla}), (\ref{pq}), (\ref{A62}) we find the interesting structure
\be
A_4^2 > {\cal P}_q > A_6^2 \,\,,
\ee
In the decompactification limit $b\to \infty$, these three quantities coincide.
Let us also consider the following relation
\be
A_6^2 =  A_4^2 - \frac{1+v^2}{(1-v^2)b^2}\,\,.
\label{Deserlike}
\ee
Whenever $A_6^2<0$, there is no horizon and it does not make sense to 
define $T_{ws}$. From (\ref{Deserlike}) we also see that it is possible to have non-zero proper acceleration $A_4$  and nevertheless the detector will remain inert. We shall provide an explicit example of this in section
\ref{sec: circular}.
Equation (\ref{Deserlike}) is reminiscent of the results
of \cite{Deser:1997ri,Deser:1998xb}, where particles
moving in $AdS$ where studied with the Unruh-De Witt formalism. 
The curvature of space-time may set a non-trivial lower bound on the proper acceleration necessary
to start feeling effects of the quantum vacuum. See also \cite{Jennings:2010vk} for recent
progress in related subjects.

It is worth mentioning that (\ref{Deserlike}) is a particular case of the general formula given in
\cite{Russo:2008gb} relating the acceleration in a curved space to the acceleration in some ambient
flat space which in our conventions reads $A_6^2 = A_4^2 + K_{uu}^2$ for
$K_{uu}^2 = - \eta_{MN}K_{uu}^M K_{uu}^N $ where
\be
K_{uu}^M = U^\alpha U^\beta (\partial_\alpha \partial_\beta l^M - \Gamma_{\alpha\beta}^\gamma
\partial_\gamma l^M) \,\,.
\ee

\section{Quarks rotating in circular paths}
\label{sec: circular}
\setcounter{equation}{0}

As advertised in the introduction, in this section we will study the problem of what happens to an accelerated observer
undergoing circular motion in a bounded space. To address this question we will use the solution  (\ref{rhoglobalsolution}) restricted to a circular motion parallel to the equator but let us first review some early results for a quark rotating in 4d Minkowski space.  

\vskip.1cm

In \cite{Athanasiou:2010pv} a quark rotating with constant angular velocity $\omega$ in Minkowski space was studied. 
The solution for the string found in that letter is a particular case of (\ref{PoincareSolution}) as we showed in section \ref{example}. We wrote the embedding functions in (\ref{Athana}).

A worldsheet horizon develops in the string worldsheet and it is located at 
$z_h=1/(\gamma^2\omega v)$ where $\gamma^2\omega v$ is the proper acceleration of the quark. 
From the perspective of an inertial observer, the quark emits gluonic radiation and this translates into stochastic fluctuation above the classical trajectory. From the point of view of the quark itself, we know that it does not see a thermal medium ({\it i.e.} there is no acceleration horizon), however, as explained in the introduction the detector clicks. This is of course consistent with the results of \cite{Padmanabhan:1983ub,Davies:1996ks}.
The authors of \cite{Athanasiou:2010pv} also calculated the total power radiated by the quark and found that,
 as for ordinary synchrotron radiation, it is proportional to the square of the proper acceleration of the
 4d motion
\be
P=\frac{\sqrt{\lambda}}{2\pi} A_4^2=\frac{\sqrt{\lambda}}{2\pi} \gamma^4 \omega^2v^2\,\,,
\label{Pliu}
\ee
in agreement with  \cite{Mikhailov:2003er}.

\subsubsection*{Quark rotating in a non-maximal circle of $S^3$}
\label{sec:parallel}

In order to pursue the ideas of \cite{Davies:1996ks} in a strong coupling situation,
we want to consider rotation of a quark in a bounded space-time and check that the
finiteness of space
 has a decisive influence on whether detectors click --- namely 
 on whether non-geodesic observers measure non-trivial excitations
associated to fluctuations of the quantum vacuum.
In the $AdS$/CFT framework, this is naturally implemented by analysing open strings
living in global $AdS$, which are dual to quarks probing a theory which lives in
$\RR \times S^3$.

\begin{figure}[htb]
\setlength{\unitlength}{1cm}
\includegraphics[width=7cm,height=7cm]{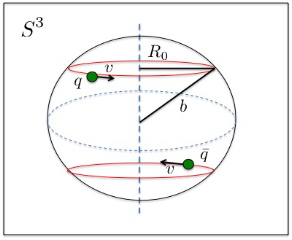}
\includegraphics[width=7cm,height=7cm]{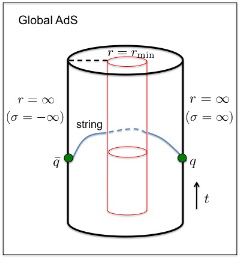}
 \begin{picture}(0,0)
\end{picture}
\caption{Left: Schematic representation of the quark and antiquark rotating in the $S^3$. The latter is located at the antipode. Right: representation of global $AdS$. The quark/endpoint is located at the boundary $r=\infty$, the string goes into the bulk, reaches a minimum value $r=r_{\text{min}}$ and comes back to the boundary. The shape of the string is just a cartoon.}
\label{sphere}
\end{figure}

In section \ref{globalAdS} we have written the general solution 
for a string moving in global $AdS$, see equation (\ref{rhoglobalsolution}) and 
remember the radius of the $S^3$ was taken to one, $b=1$. 
We now want to consider the particular case in which the string endpoint is moving with constant
angular velocity along a non-maximal circle of the $S^3$ (see Figure \ref{sphere}). 
The first step is to give an explicit representation of the $n^a$ defined in section 
\ref{globalAdS} in terms of a set of coordinates parameterizing the $S^3$. We write\footnote{
In order to make more obvious the relation to the Poincar\'e patch case, we will
be a bit cavalier with notation. The coordinates (\ref{n1n4}) are not good for the whole $S^3$.
Strictly speaking, instead of $R$, 
one should use another angle $R=\sin \chi$ with $\chi \in [0,\pi]$.
It should be understood in the following that the string lies at $\chi \in [0,\pi/2]$
for $\sigma \geq 0$ and at $\chi \in [\pi/2, \pi]$ for $\sigma \leq 0$.}
\bear
&&n^1 =  \sqrt{1-R^2}\,\,,\qquad\qquad\quad
n^2 = R\cos \theta\,\,,\rc
&&n^3 =R \sin\theta \cos\phi\,\,,\qquad
n^4 = R\sin\theta\sin\phi\,\,.
\label{n1n4}
\eear
with $R\in [0,1]$, $\theta \in [0,\pi)$ and $\phi \in [0,2\pi)$. The metric of the unit $S^3$ then reads
\be
d\Omega_3^2 = \frac{dR^2}{1-R^2} + R^2 d\theta^2 + R^2\sin\theta d\phi^2\,\,.
\ee
Let us now consider the string endpoint to be in circular motion with
constant velocity in the azimuthal angle $\phi$ (see Figure \ref{sphere}):
\be
\bar R(\tau) =R_0 \,\,,\qquad
\bar \phi(\tau)= \omega \, t_{ret}(\tau)\,\,,\qquad
\bar \theta(\tau) = \frac{\pi}{2}
\ee
where $R_0$ is the radius of the trajectory at the boundary.
We can read the quantities defined in (\ref{vaaadefs}), (\ref{v2a2defs}):
\be
v^2 = R_0^2 \omega^2 \,\,,\qquad
\vec v \cdot \vec a = 0 \,\,,\qquad
a^2 = R_0^2 \omega^4\,\,.
\label{v2a2rot}
\ee
From (\ref{globalttnn}) we also know that
\be
\tret = \frac{\tau}{\sqrt{1-\omega^2 R^2_0 }}=\gamma \tau\,\,.
\ee
Restricting (\ref{rhoglobalsolution}) to this rotating quark ansatz, 
the string embedding is given by
\bear\label{embedding}
t(\tau, \sigma) &=& \gamma \tau + \arctan \left(\frac{\gamma}{\sigma}\right)\,\,,\rc
r(\tau, \sigma) &=& \sqrt{\sigma^2 + \gamma^2 \omega^2R^2_0}\,\,,\rc
\theta(\tau,\sigma) &=& \frac{\pi}{2} \,\,,\rc
\phi (\tau,\sigma) &=& \gamma \omega \tau+ \arctan \left(\frac{\gamma \omega}{\sigma}\right)\,\,,\rc
R(\tau, \sigma) &=& R_0\sqrt{\frac{\sigma^2+\gamma^2\omega^2}{\sigma^2+\gamma^2\omega^2R^2_0}}\,\,.
\eear

The solution written in this form is not very illuminating. We can eliminate the first two equations in (\ref{embedding}) by rewriting the solution just as a function of $t$ and $r$. The result is
\be\label{eqR}
R(r)=R_0 \sqrt{\frac{1+ \gamma^2\omega^2}{r^2}(1-R^2_0)}
\ee
and 
\be\label{eqfi}
\phi(t,r)=\omega t- \omega\arctan({\frac{\gamma}{\sqrt{1+r^2-\gamma^2}}})+\arctan({\frac{\omega\gamma}{\sqrt{1+r^2-\gamma^2}}})\,\,.
\ee
Reinserting the $S^3$ radius $b$ and taking the appropriate decompactification limit, it
is not difficult to check that (\ref{eqR}), (\ref{eqfi}) reduce to (\ref{Athana}).
As expected, when $\sigma \rightarrow \infty$ then $R(r)\rightarrow R_0$ and $\phi(t,r)\rightarrow \omega t$.
Notice also that the embedding functions are not well defined for all values of the coordinate $r$.
From (\ref{eqfi}) it is straightforward to see
 that $r\in(r_{\text{min}},\infty)$ where $r_{\text{min}}=\gamma\omega R_0$,
 which corresponds to $\sigma=0$ (see Figure \ref{sphere}).

Plugging (\ref{v2a2rot}) into (\ref{pq}), (\ref{PVilla}),  (\ref{A62}), we can read the quantities:
\be
{\cal P}_q= \gamma^4 R_0^2 \omega^2 (\omega^2 -1) \,\,,\quad
A_4^2= \gamma^4  R_0^2 \omega^4 (1-R_0^2)\,\,,\quad
A_6^2= \gamma^4  (R_0^2 \omega^4 - 1)\,\,.
\label{rot}
\ee

Let us now try to extract some physical information from this solution.
First, from the induced metric we can obtain the position of the worldsheet horizon. The result is
\be\label{wshorizon}
r_H=\sqrt{\gamma^4R^2_0\omega^2(\omega^2-1)-1}\,\,.
\ee
We are now dealing with a quark (detector) rotating in a bounded space so we expect to make contact with the results of \cite{Davies:1996ks}. It is clear from (\ref{wshorizon}) that not for every value of $\omega$ and $R_0$ the string worldsheet develops a horizon. One way to see this is to fix the value of $R_0$ (i.e. fix the
``latitude" at which the string endpoint is moving) and from (\ref{wshorizon}) 
find the critical value  $\omega=\omega_c$ for which a horizon appears. This happens when
$r_H = r_{min}$, leading to
\be
\omega_c=\frac{1}{\sqrt{R_0}}\,\,.
\ee 
This is the same condition as demanding $A_6^2=0$.
Of course the endpoint cannot exceed the speed of light, then we can conclude that the observer will feel a non-trivial interaction with the quantum vacuum for a range of angular velocities:
\be
\frac{1}{\sqrt{R_0}} < \omega < \frac{1}{R_0}
\ee
This is in qualitative agreement with what happens in the weakly coupled case
 analyzed by Davies et al. \cite{Davies:1996ks}

It is also interesting to look at the value of the rate of energy flowing down the string from 
the quark to the antiquark ${\cal P}_q$, 
see (\ref{rot}). It turns out that at $\omega=1$, this quantity changes sign.
This is precisely the value of $\omega$ at which a co-rotating observer in the equator moves at the
speed of light, although we do not know whether there is any physical reason for this coincidence. 
Thus, there are  three different physical regimes.
Reinserting $b$ (the radius of the $S^3$) they are:
\begin{itemize}

\item $0 < \omega\,b < 1$ 

There is no worldsheet horizon.
A co-rotating observer at the equator follows a time-like trajectory.
Moreover,
the quark is not loosing energy (${\cal P}_q < 0$), but rather
 gaining energy from the gluonic fields whereas
 the anti-quark located at the antipode does emit energy. 
 Below we will show that the external force acting on the quark is also negative, meaning that in order to have a stationary motion we need to pullback the quark instead of pushing on it. This makes it clear that in this situation, one should
 definitely not consider the quark and antiquark as independent objects, but one should rather think of the 
whole system as a kind collective motion. Both finite size and strong coupling seem to play a role
in order to make this possible.
The particular case $\omega\,b=1$ implies that ${\cal P}_q=0$ (and also the external force
vanishes), then, the quark and antiquark rotate without any energy flow along the string. 

\item $1 < \omega\,b < \frac{1}{\sqrt{R_0}}$

A would-be co-rotating observer at the equator would now move faster than the local speed of light.
 The total power radiated
 ${\cal P}_q$ is positive (and also the external force). Nevertheless, there is not a worldsheet horizon and according to our proposal this means that the detector is not interacting with the quantum vacuum. As explained in the introduction, a similar situation is observed in \cite{Davies:1996ks}. 

\item  $\frac{1}{\sqrt{R_0}} < \omega\,b < \frac{1}{R_0}$

The presence of the worldsheet horizon indicates, from the point of view of the inertial observer, that the quark will fluctuate around its classical trajectory (stochastic motion), meaning that the detector is interacting non-trivially with the vacuum. It should be interesting to describe the stochastic motion of the endpoint/quark following the ideas of \cite{deBoer:2008gu,sonteaney} but we leave that for future work. In this regime, the quark is emitting gluonic radiation which, as explained before, lives for a finite amount of time before it reaches the antiquark and is absorbed.
\end{itemize}

Let us make a heuristic argument in order to explain the transition between the second and
third regimes. 
 Let us think of the quark  in a trajectory near the pole of the sphere such that it does not
really probe the finiteness of the space. In the limit $b \to \infty$, $R_0 \to 0$ with $b\, R_0$ fixed,
the situation is essentially as
if the quark was rotating in Minkowski space \cite{Athanasiou:2010pv},
see section \ref{example}. Only the third regime exists in this limit and there must be a worldsheet horizon.
Synchrotron radiation is emitted and the length-scale controlling the actual size of a pulse
is $\gamma^{-3} a^{-1}$  \cite{Athanasiou:2010pv}\footnote{This observation was suggested
in \cite{Athanasiou:2010pv} but not proved in general, see also \cite{Hubeny:2010rd}
for a different approach to the same issue.}. 
The idea is that when the size of this pulse
is of the same order as the radius of the $S^3$, namely $\gamma^{-3} a^{-1} \sim b$, one cannot think any
more of a quantum process in which a gluon emitted at a point and absorbed at the antipode.
In a sense, the wavelength of the emitted ``gluon'' is as large as the $S^3$. At this point, the stochastic
process of backreaction to the radiation emission ceases to exist. If we go to 
$\omega\,b \sim \frac{1}{\sqrt{R_0}}$, we would have $\gamma$ of order one and indeed
$\gamma^{-3} a^{-1} \sim b$. This could give a qualitative understanding of the 
$\omega_c \sim 1/\sqrt{R_0}$ behaviour. In order to make this connection more rigorous, one
should compute, along the lines of \cite{Athanasiou:2010pv}, the profile of radiation for the present
situation.

\subsubsection*{Force and momentum}

As a consistency check we can calculate the total power radiated by the rotating quark but applying the techniques used in \cite{Athanasiou:2010pv}. 
The external force acting on the string endpoint is given by
\be
\mathcal{F}_{\phi}=\frac{\sqrt{\lambda}}{2\pi} \Pi^\sigma_{\phi}|_{\sigma=\infty}=-\frac{\sqrt{\lambda}}{2\pi} \frac{\partial\mathcal{L}}{\partial\phi'}
\ee
where $'$ means $\partial/\partial \sigma$. 
For this particular case
\be
\frac{\sqrt{\lambda}}{2\pi} \Pi^\sigma_{\phi}|_{\sigma=\infty}=\frac{\sqrt{\lambda}}{2\pi} \Pi=\frac{\sqrt{\lambda}}{2\pi}\gamma^4R^2_0\omega(\omega^2-1)\,\,.
\ee
Finally, using the fact that energy is conserved on the string worldsheet we can write
\be
\omega\,\mathcal{F}_{\phi}
=\frac{\sqrt{\lambda}}{2\pi} \gamma^4R^2_0\omega^2(\omega^2-1)
\ee
which is exactly $\frac{\sqrt{\lambda}}{2\pi} {\cal P}_q$, with ${\cal P}_q$ given in
(\ref{rot}). The result in the Poincar\'e patch (\ref{Pliu}) is recovered neglecting the 1 inside
the bracket.
For the anti-quark endpoint, one gets the same result with opposite sign.

\appendix

\section*{APPENDICES}

\section{Global embedding in Minkowski space}
\label{sec:GEMS}
\setcounter{equation}{0}

In section \ref{sec: general} we checked that, 
whenever $A_6^2$ is constant and positive, there is a worldsheet horizon and
its temperature can be associated to what one 
could loosely call the Unruh temperature,
see (\ref{unruhgen}). 
This poses a puzzle. On the one hand, we
would only expect uniformly accelerated motion to precisely produce 
 thermal effects. But we have found that any stationary
motion with the same constant modulus of the proper acceleration
produces exactly the same induced metric on the worldsheet and therefore,
it will yield the same pattern of stochastic motion --- at least along the directions transverse to the
classical trajectory, including excitations along the internal space.
There is no difference between uniform acceleration and other stationary motions, from this point of view.
However, this is not a prove that the quark is, in general, immersed in a thermal bath.
In this appendix, we will make a computation that singles out uniform acceleration: we will check that
only in that case, every bit of the string worldsheet is locally immersed in a thermal bath.

Let us start by considering the local temperature as seen from any bit of the string worldsheet, at some
given $\sigma$. That is, we compute
a temperature depending on $\sigma$, making use of Tolman's law for the worldsheet metric
(\ref{dsws2again})
\be
\hat T_{ws}(\sigma)=\frac{T_{ws}}{\sqrt{g_{\tilde\tau \tilde \tau}}}=\frac{T_{ws}}
{\sqrt{\sigma^2 - A_6^2}}\,\,.
\label{Twss}
\ee
Here and in the following, we use hats for quantities referring to the string bit located
at a given $\sigma$.

Before proceeding further, it is useful to introduce notation for the higher derivatives of the position.
The relativistic jerk and snap defined in \cite{Russo:2009yd} are, for a particle living in
$n$-dimensional Minkowski space time with proper time $\tau$, $n$-velocity
$U^\mu$ and $n$-acceleration $A^\mu = \partial_\tau U^{\mu}$:
\be
\Sigma^\mu = \frac{dA^\mu}{d\tau} + (A^\nu A_\nu) U^\mu\,\,,\qquad
\Xi^\mu = \frac{d\Sigma^\mu}{d\tau} + (\Sigma^\nu A_\nu) U^\mu\,\,,
\label{defSigma}
\ee
such that $U^\nu \Sigma_\nu = U^\nu \Xi_\nu=0$. A particle in uniform acceleration has vanishing
relativistic jerk $\Sigma^\mu$ and snap $\Xi^\mu$.

We can now compute a different temperature for the same string bit: we consider the string bit
itself as a particle and compute the temperature it feels due to its motion within the $AdS_5$
space.
In order to do that, we follow ideas of \cite{Deser:1997ri,Deser:1998xb}
and find out how it moves within the GEMS (global embedding in Minkowski space) of $AdS$.
The GEMS of $AdS$ is simply given by the $X^M$ of  (\ref{hyperboloid}) and therefore
we already have the expression for the motion in the GEMS written in (\ref{vofl}).
All of the following computations follow from this solution.

 The 
proper time corresponding to the string bit is
\be
\hat \tau (\sigma) = \tau \sqrt{\sigma^2 - A_6^2}
\ee
such that
\be
\hat U^M(\sigma) = \frac{dX^M}{d\hat \tau}= \frac{A^M + \sigma  U^M}{\sqrt{\sigma^2 - A_6^2}}
\label{bulkU}
\ee
where we have introduced 
\be
 U^M = \partial_\tau l^M\,\,,\qquad
 A^M = \partial_\tau U^M\,\,.
\ee
These quantities without hat refer to the motion of the string endpoint
within the 6d space spanned by the $l^M$.
Remember that $U^M U_M=1$, $A^M A_M = -A_6^2$.
Let us also define $- \Sigma^M \Sigma_M \equiv  \varsigma^2 \,\,,\
- \Xi^M \Xi_M \equiv   \xi^2$. For simplicity, we assume that $A_6^2$, $\varsigma^2$,
$\xi^2$ are all constant as is the case for circular motion.
The acceleration of the string bit 
at position $\sigma$ is $\hat A^M(\sigma) =\frac{d\hat U^M}{d\hat \tau}$.
It is straightforward to compute its modulus
\be
\hat A^2(\sigma) = - \hat A^M \hat A_M = \frac{A_6^2}{\sigma^2 - A_6^2}+\frac{\varsigma^2}{(\sigma^2 - A_6^2)^2}\,\,.
\label{bulkA2}
\ee
Notice that in the right hand side, $A_6^2$ and $\varsigma^2$ refer to quantities associated
to the motion of the string endpoint.
Imagine we define an Unruh-like
 ``acceleration temperature" as $\hat T_A (\sigma) = \frac{\hat A}{2\pi}$. Then, comparing to
 (\ref{Twss}), we find that the local worldsheet temperature coincides with the acceleration temperature
only when the boundary jerk vanishes, namely for uniform acceleration (a case studied in
\cite{Paredes:2008cr}). 
This is suggesting that the nature of the worldsheet is only
really thermal in that case, as one would expect on general grounds.

On the other hand, the quantity $\hat T_A(\sigma)$ can really be considered as a temperature 
only if the relativistic jerk is small for the corresponding string bit 
\cite{Russo:2008gb}\footnote{In fact, all higher derivatives of the position should be small
 in order for a motion to be approximately in uniform acceleration
and therefore for $\hat T_A$ to be a temperature \cite{Russo:2009yd}.}.
The appropriate dimensionless quantity to be used as a measure of how small the jerk is
was
 defined in \cite{Russo:2008gb,Russo:2009yd}. We find
\be
(\hat \lambda^{(1)})^2= \left| \frac{\hat \Sigma^M
\hat \Sigma_M}{ (\hat A^N \hat A_N)^2}\right|^2
= \frac{
(\sigma^2 - A_6^2)(\xi^2 + \sigma^2 \varsigma^2)+ \varsigma^4
}{(A_6^2(\sigma^2 - A_6^2)+\varsigma^2)^2
}\,\,.
\label{lambda12}
\ee
If the jerk and snap of the boundary trajectory vanish $\varsigma^2=
\xi^2=0
$, this quantity is identically zero.
Otherwise, and provided
 there is a horizon ($A_6^2>0$), then
we  have $(\lambda^{(1)})^2=1$ precisely at
$\sigma^2 =A_6^2$. Thus, unless the quark is uniformly
acclerated, it is not possible to have a small
jerk for the whole string.  This seems to point out that the string only is really
in a thermal bath when the quark undergoes uniform acceleration. This is only
possible when the field theory is defined in $\RR^{1,3}$, and not in $\RR \times S^3$.

\section{A non-singular coordinate transformation}
\label{app:coordchange}
\setcounter{equation}{0}

As explained in the introduction, there are at least
three different ways to understand the detection of particles from the point of view of an accelerated observer and all of them are equivalent in the uniformly accelerated case. For more general motions this is no longer true.

In this appendix we give the explicit coordinate transformation that relates the inertial observer to the accelerated one (sitting on the quark). We show that the transformation is non-singular and therefore the determinant of the jacobian matrix $\mathcal{J}(r,t,\phi,R,\theta)$ is different from zero.
Given the embedding fuctions (\ref{eqR}) and  (\ref{eqfi}), the coordinate transformation is
\bear\label{coortrans}
t'&=&t\,\,,\rc
r'&=&r\,\,,\rc
\theta'&=&\theta\,\,,\rc
\phi'&=&\phi-\omega t+ \omega\arctan({\frac{\gamma}{\sqrt{1+r^2-\gamma^2}}})-\arctan({\frac{\omega\gamma}{\sqrt{1+r^2-\gamma^2}}})\,\,,\rc
R'&=&R-\sqrt{\frac{\gamma^2\omega^2R^2_0}{r^2}(1-R^2_0)+R^2_0}\,\,.
\eear
which is non-singular everywhere.
Using (\ref{coortrans}) we can calculate de determinant of the jacobian matrix, the result is
\be\label{determinant}
\det[{\mathcal{J}(r,t,\phi,R,\theta)}]=1
\ee
reflecting the fact that the Bogolubov transformation is trivial \cite{Letaw:1979wy} or equivalently, that there is not an acceleration horizon present in the transformed metric.
This is very different for example to what happens when using the Rindler coordinate system to go from Minkowski to Rindler space-time for which the $\det[{\mathcal{J}}]=0$ precisely at the location of the acceleration horizon.

\section*{Acknowledgements}

We are especially indebted to J. Russo for enlightening discussion and collaboration 
during early stages of this work.
We also want to thank R. Emparan, B. Fiol, J. Garriga, A. G\"uijosa, D. Mateos 
and P. K. Townsend for very useful conversations.

The research of A.P and M.C is
supported by grants FPA2007-66665C02-02 and DURSI
2009 SGR 168, and by the CPAN CSD2007-00042 project of the
Consolider-Ingenio
2010 program. The research of M.C is also supported by a postdoctoral fellowship from Mexico's National Council of Science and Technology (CONACyT).

\end{document}